\documentclass[aps,prd,nofootinbib,showpacs,amssymb]{revtex4}

\usepackage{amsmath,amsfonts}

\newcommand{\be}{\begin{equation}}
\newcommand{\ee}{\end{equation}}
\newcommand{\bea}{\begin{eqnarray}}
\newcommand{\eea}{\end{eqnarray}}

\begin{document}

\author{David Brizuela}
\email{D.Brizuela@uni-jena.de}

\author{Gerhard Sch\"afer}
\email{G.Schaefer@tpi.uni-jena.de}

\affiliation{Theoretisch-Physikalisches Institut,
Friedrich-Schiller-Universit\"at, Max-Wien-Platz 1, 07743 Jena, Germany}

\title{Fourth-post-Newtonian-exact approximation to General Relativity}

\date{\today}

\begin{abstract}
An approximation to General Relativity is presented that
agrees with the Einstein field equations up to and including the
fourth post-Newtonian (PN) order. This approximation is formulated
in a fully constrained scheme: all involved equations
are explicitly elliptic except the wave equation that
describes the two independent degrees of freedom of the gravitational
field. The formalism covers naturally the conformal-flat-condition (CFC)
approach by Isenberg, Wilson, and Mathews and the
improved second PN-order exact approach CFC+.
For stationary configurations, like Kerr black holes,
agreement with General Relativity is achieved even
through 5PN order. In addition, a particularly
interesting 2PN-exact waveless approximation is analyzed in detail,
which results from imposing more restrictive conditions.
The proposed scheme can be considered as a further
development on the waveless approach suggested by Sch\"afer and
Gopakumar [Phys. Rev. D {\bf 69}, 021501 (2004)].
\end{abstract}

\pacs{04.20.Ex, 04.25.Nx, 04.30.Db}

\maketitle

\section{Introduction}

The solution of the Einstein field equations for inspiraling compact
binaries is one of the great challenges of Numerical Relativity, e.g., \cite{C00}.
To fully succeed with this problem realistic initial data are needed
which do not include incoming (spurious, ``junk'') gravitational waves.
If one could follow up the evolution from the very beginning of the
inspiraling process, i.e., right from the infinite separation of the
both objects and without gravitational waves present at this initial stage,
the configuration to start with at any finite instant of time would be well under
control just by calculating the history up to that instant of time.
Lack of this piece of information requires the construction of initial
value data by other means.

On the other hand, there are several astrophysically relevant scenarios,
like supernovae explosions or initial stages of a binary black hole,
which can be described very accurately with approximated versions
of Einstein equations. These two issues (initial data generation and
construction of truncated schemes of General Relativity) are deeply
connected since by extending to all times the conditions imposed
on the freely specifiable part of the initial data set, one can
obtain an approximated version of Einstein equations.

As is well known, in the initial slice, we are free to choose
four gravitational degrees of freedom from the initial data set
(the two independent field degrees of freedom and their time derivatives
or, their canonically conjugate momenta), whereas the remainder
gravitational objects must be obtained through the resolution of the constraint equations
and coordinate conditions.
Resorting to a canonical formulation
of General Relativity, e.g., the Arnowitt-Deser-Misner (ADM) one
\cite{ADM62} -- to be used throughout in this paper --,
these independent gravitational degrees of freedom get clearly identified.
The four degrees of freedom are encoded in
the two independent metric-field components $h^{\rm TT}_{ij}$ and their
canonical conjugate $\frac{c^3}{16\pi G}\pi_{\rm TT}^{ij}$, where $\rm TT$ means transverse
and traceless with respect to an auxiliary 3-dimensional flat metric $\delta_{ij}$ with $i,j =
1,2,3$. Therefore, the initial value problem reduces to the fixation of the spatial
dependence of these two objects at a finite initial time.

In the Isenberg-Wilson-Mathews conformal-flat condition
(CFC) approach to General Relativity \cite{I78,WM89}, $h^{\rm TT}_{ij}$
and its time derivative $h^{\rm TT}_{ij,0}$ are chosen to be vanishing.
Hence, this scheme is very appropriate to construct conformally flat initial
data. On the other side, understood as an approximation to General Relativity
so that the mentioned conditions ($h_{ij}^{\rm TT}=h_{ij,0}^{\rm TT}=0$)
are valid for all times, this formalism reproduces the evolution of
the gravitating system at 1PN order,
which implies deviations from General Relativity at 2PN.

Even though, near-zone post-Newtonian (PN) calculations show
that for inspiraling binaries with no-incoming radiation, $h^{\rm TT}_{ij}$ can never be exactly
zero, see, e.g., \cite{OO74,S85,JS97,JS98,NB05}, but contain conservative and dissipative
terms which destroy CFC. The first piece of deviation from CFC appears at 2PN order
and is of conservative type. The higher-order pieces appearing in the cited
papers are 2.5PN (dissipative), 3PN (conservative), and 3.5PN
(dissipative). Therefore, many attempts have been made to construct
conformally nonflat schemes, see for instance \cite{BCT98, Lag99}.
In particular, regarding works done based on PN corrections,
the so-called CFC+ approach, as detailed in \cite{CD05}, generalizes
the CFC framework by just correcting for the additional 2PN piece. In the
scheme \cite{TB03} also 2.5PN terms have been considered and were recently
further developed in \cite{KT07,KT09}. The 2.5PN and 3.5PN pieces have
been added within the CFC-type skeleton approach \cite{FJS04} in
\cite{GoSc08}. However, in this last reference the radiation reaction has been
taken into account even through 6PN order. Herein, the 4PN, 5PN, and 6PN orders
are both conservative and dissipative, where the dissipative parts result
from the peculiar tail structure of the radiation process.

The separation in dissipative and conservative terms allows the
separate calculation of the dissipative and conservative parts of the metric coefficients.
The aim of the waveless approximations to General Relativity beyond CFC
is the determination of conservative terms in the gravitational
radiation process. Related with waveless approaches are
the near-zone helically symmetric schemes introduced by \cite{BD92},
which also do not allow for emission of waves and have been intensively
studied in the recent past, e.g., \cite{SU04,UL06,UL09}.
The fully constraint formulation \cite{BG04} has found deeper
investigations and waveless applications in \cite{CC08,CC09}.
A comparison between helically symmetric and waveless description
of binary systems has been performed in, e.g., \cite{YBR06}.

In the present paper, a modification of the waveless approach by \cite{ScGo04} is developed
which supplies a treatment of radiating systems in agreement with full General Relativity
through 4PN order and even up to 5PN order under nonradiating conditions.
The CFC and CFC+ approaches are well covered as well as
treatments including outgoing radiation \cite{TB03,KT07,KT09,JY09}.
This approximation is formulated in a fully constrained scheme.
All involved equations are explicitly elliptic except the
wave equation obeyed by $h_{ij}^{\rm TT}$ that describes
the two degrees of freedom of the gravitational field.
This hyperbolic equation must be fed with freely specifiable
initial data $h_{ij}^{\rm TT}$ and $h_{ij,0}^{\rm TT}$.

In addition, we analyze in detail a particularly
interesting 2PN-exact waveless approximation to General
Relativity, which is a particular case of our 4PN approach.
More precisely, it results from imposing the ``conjugate''
conditions to CFC, namely $\pi^{ij}_{\rm TT}=\pi^{ij}_{\rm TT}{}_{,0}=0$.

Under stationary conditions only integer PN orders are different from
zero. We will keep to this integer PN-counting also under nonwaveless
conditions where half-integer orders appear. This particularly means
that our proposed 4PN-exact approach is, in precise terms, a
4.5PN-exact approach as well.

The article is organized as follows. In Sec. \ref{ADM} the Einstein
field equations are shown in the context of the ADM formalism.
Section \ref{Elliptic} presents explicit elliptic equations that
are valid in full General Relativity and allows one to solve
for all the geometric objects but the transverse
and traceless part of the metric and of its conjugate momentum.
In Sec. \ref{2PN} the transverse and traceless part of the conjugate
momentum is assumed to be vanishing, which leads to a 2PN-exact
waveless approximation scheme. Section \ref{4PN} generalizes the
previous scheme to a 4PN-exact framework suggesting another 2PN-exact
waveless approach. We show that this formalism
reduces to the well-known CFC+ approach under the corresponding
conditions in Sec. \ref{CFC}. The particular matter model
of a perfect fluid is presented in Sec. \ref{Matter}. Finally
the conclusions are drawn in Sec. \ref{Conclusions}.

\section{Einstein field equations in ADM formalism}\label{ADM}

In the ADM formalism of General Relativity, the spacetime line element
is split into the $(3+1)$ form,
\be
ds^2 = - \alpha^2 c^2 dt^2 + \gamma_{ij}(dx^i + \beta^icdt)(dx^j+\beta^jcdt),
\ee
where $\alpha$ is the lapse function, $\beta^i$ the shift vector,
$\gamma_{ij}$ the induced metric on a three-dimensional 
spatial slice $\Sigma (t)$, parameterized by the
time coordinate $t$, and $c$ is the speed of
light. The three-metric $\gamma_{ij}$ and its canonical conjugate 
$ \frac {c^3 }{ 16\,\pi\,G}\,\pi^{ij}$, which is 
a contravariant symmetric tensor
density of weight $+1$, satisfy  the Hamiltonian and  momentum constraints
\cite{ADM62},
\be
\gamma^{1/2} \mbox{R} = \frac{1}{2\gamma^{1/2}}(2\pi^{ij} \pi_{ij}-\pi^2)
+\kappa \gamma^{1/2} \alpha^2 T^{00}\,,
\label{Eq.2.2}
\ee
\be
(-2 \pi_i{}^j{}_{|j}=) -2 \pi_i{}^j{}_{,j} + \pi^{kl} \gamma_{kl,i} = \kappa \gamma^{1/2}  \alpha T^0{}_i\,,
\label{Eq.2.3}
\ee
where  $\mbox{R}$ is 
the curvature scalar of $\Sigma (t)$,
$\gamma$ the determinant of $\gamma_{ij}$,
$\pi^i{}_j = \gamma_{jk}\pi^{ik}$, and $\pi = \pi^i{}_i$.
On the other hand,
$T^{00}$ and $ T^0{}_i$ are the components of
the unspecified  four-dimensional stress-energy tensor for the matter
source $T^{\mu\nu}$.
The canonical conjugate $\pi^{ij} $ is related to $K_{ij}$, the extrinsic
curvature of $\Sigma(t)$, by 
$\pi^{ij} = - \gamma^{1/2}
(\gamma^{il} \gamma^{jm} - \gamma^{ij} \gamma^{lm}) K_{lm}$,
where $ \gamma^{ij} $ is the inverse metric of $\gamma_{ij}$.
In the above equations,
a partial derivative is denoted by a comma, whereas $|$
stands for the three-dimensional covariant derivative,
and $\kappa \equiv \frac{16\pi G}{c^4}$, with $G$ the Newtonian gravitational constant.
The three-metric and its canonical conjugate evolve in accordance with
the following evolution Eqs. \cite{ADM62},
\bea
\pi^{ij}_{~~,0} &=& -\frac{1}{2}\alpha \gamma^{1/2} (2\mbox{R}^{ij} - \gamma^{ij}\mbox{R}) 
+ \frac{1}{4}\alpha\gamma^{-1/2}\gamma^{ij}(2\pi^{kl} \pi_{kl}-\pi^2)
-\alpha \gamma^{-1/2}(2\pi^{ik} \pi^j{}_k-\pi\pi^{ij})
\nonumber \\
&+&  \gamma^{1/2}(\alpha^{|ij}-\gamma^{ij}\alpha_{|k}{}^{k})
+ (\pi^{ij}\beta^k)_{|k} - \pi^{kj} \beta^i_{~|k} - \pi^{ik} \beta^j_{~|k}
+ \frac{\kappa}{2} ~ \alpha \gamma^{1/2} \gamma^{ik}\gamma^{jl} T_{kl},
\label{Eq.2.4}
\eea
and 
\be
\gamma_{ij,0} = \alpha \gamma^{-1/2}(2\pi_{ij} - \gamma_{ij}\pi) + \beta_{i|j} +
\beta_{j|i},
\label{Eq.2.5}
\ee
where $\mbox{R}_{ij}$ is the Ricci
tensor associated with $\Sigma (t)$.
In this paper, we raise and lower indices of three-dimensional
objects with $\gamma^{ij}$ and  $\gamma_{ij}$ respectively.

The ADM coordinate conditions, which generalize the isotropic
Schwarzschild metric, read
\be
\pi^{ii}=0,
\label{Eq.2.6}
\ee
where, and from here onwards, repeated covariant or contravariant
indices imply the usage of Einstein summation convention, and
\be
\gamma_{ij}= \Psi\,\delta_{ij}+h^{\rm TT}_{ ij},
\label{Eq.2.7}
\ee
where $\Psi$ is a conformal scalar and $h^{TT}_{ij}$ 
the transverse and traceless part of the
three-metric $\gamma_{ij}$ with respect to the Euclidean 3-metric
$\delta_{ij}$. By definition, $h^{TT}_{ij}$ satisfies 
$h^{\rm TT}_{ii} = h^{\rm TT}_{ ij,j} = 0$.
The conformally-flat condition $h^{\rm TT}_{ij} = 0$ gives 
a simple expression for the three-dimensional curvature scalar,
$\gamma^{1/2} \mbox{R} = - 8\,\Psi^{1/4}\,\Delta \Psi^{1/4}$,
where $\Delta$ stands for the three-dimensional Euclidean Laplacian.
The gauge fixing equation for $\gamma_{ij}$ (\ref{Eq.2.7})
can be rewritten in a differential way,
\be
3 \gamma_{ij,j} - \gamma_{jj,i} = 0.
\label{Eq.2.8}
\ee

Equation (\ref{Eq.2.6}) results in the covariant trace of $\pi^{ij}$ of the
form $\pi =\pi^{ij} h^{\rm TT}_{ij}$. Taking into account the
space-asymptotic properties  $\pi^{ij} \sim 1/r^2$ and $h^{\rm TT}_{ij} \sim
1/r$, the gauge condition (\ref{Eq.2.6}) turns out to mean asymptotic maximal
slicing ($K\sim 1/r^3$). The gauge conditions (\ref{Eq.2.6}) and (\ref{Eq.2.7}),
or (\ref{Eq.2.8}), are very close to the well-known Dirac gauge
conditions which include maximal slicing. 
In fact, condition (\ref{Eq.2.8}) is identical with the corresponding
Dirac gauge condition to linear order in $\gamma_{ij} - \delta_{ij}$.

As will be made explicit in the next section,
the function $\Psi$, with $\Psi - 1  \sim 1/r$ at asymptotic infinity,
and the longitudinal part of the momentum ${\pi}^{ij}$ are determined using the
Hamiltonian and momentum constraints, given in Eqs. (\ref{Eq.2.2})
and (\ref{Eq.2.3}),  by elliptic equations. On the other hand,
the elliptic equations for the lapse $\alpha$, with $\alpha -1  \sim
1/r$, and the shift $\beta^i$, with $\beta^i\sim 1/r$, result from the evolution Eqs. (\ref{Eq.2.4})
and (\ref{Eq.2.5}), respectively, after applying the coordinate conditions
(\ref{Eq.2.6}) and (\ref{Eq.2.8}). Since we want to provide a set of
equations that can be numerically solved, we also will take care
that all the sources decay at least as $1/r^4$ at asymptotic
infinity, avoiding in this way convergence problems.

Finally, it is important to stress that our variables are
those dictated by a canonical formulation. This fact clearly
makes them form a fully consistent set of independent variables.
Of course, other sets of variables may also be chosen, like the
one presented in Ref. \cite{BG04} or the
Kol-Smolkin one  \cite{KoSm08}, which is connected with the Landau-Lifshitz
decomposition of the metric \cite{LL} and has
extensively been used in, e.g., \cite{GiRo08}. Even though,
it should be pointed out that only within a radiation-type (``Coulomb-type'')
set of coordinate conditions those variables fulfill a mixed
set of elliptic and evolutionary (hyperbolic) field equations
instead, e.g., a purely hyperbolic system.

\section{Explicit elliptic equations}\label{Elliptic}

In this section we will present explicit elliptic equations to solve
for all the objects but the transverse and traceless parts of the
three-metric $\gamma_{ij}$ and of its corresponding momentum
$\pi^{ij}$. No assumptions will be made, hence these equations will
be valid in full General Relativity.

\subsection{Equation for the shift}

Taking the time derivative of condition (\ref{Eq.2.8}) and making use of
Eq. (\ref{Eq.2.5}), we obtain an elliptic equation for the shift,
\begin{equation}
(\beta_{i|j})_{,j} + (\beta_{j|i})_{,j} - \frac{2}{3} (\beta_{j|j})_{,i}
= - (\alpha \gamma^{-1/2} (2\pi_{ij} - \gamma_{ij}\pi))_{,j} 
+ \frac{1}{3}(\alpha \gamma^{-1/2} (2\pi_{jj} - \gamma_{jj}\pi))_{,i}.
\end{equation}
It can be more explicitly written down in the following way,
\begin{eqnarray}
\Delta\beta_i +\frac{1}{3}\beta_{j,ji}
&=&
-\frac{2}{3}(\gamma^{jk}\beta_j \Gamma_{kll})_{,i}
+2 (\gamma^{jk}\beta_j \Gamma_{kil})_{,l}+ (\alpha \gamma^{-1/2} \gamma_{ij}\pi)_{,j}
+ \frac{1}{3}(\alpha \gamma^{-1/2} (2\pi_{jj} - \gamma_{jj}\pi))_{,i}
\nonumber\\\label{betaeq}
&-&2 \pi_{ij}(\alpha\gamma^{-1/2})_{,j}\!-\frac{2\alpha}{\gamma^{1/2}}
\left[
\pi^j{}_i\Psi_{,j}+h^{\rm TT}_{jk}\,\pi^{j}{}_{i,k}+\frac{\Psi}{2}
\left(
\pi^{kl}h^{\rm TT}_{kl,i}-\kappa\gamma^{1/2}\alpha T^0{}_i
\right)
\right],
\end{eqnarray}
where we have applied the momentum constraint (\ref{Eq.2.3}) to replace the
term $\pi_{ij,j}$ in order to guarantee an adequate fall-off behavior
($\sim 1/r^4$) at asymptotic infinity of the right-hand side of this equation.
Here one should use the usual definition for Christoffel symbols,
\begin{equation}
\Gamma_{ijk} = \frac{1}{2}(\gamma_{ij,k}+\gamma_{ki,j}-\gamma_{jk,i}),
\end{equation}
and the exact relations \cite{S85},
\begin{eqnarray}
\gamma\gamma^{ij}&=&\chi\delta_{ij} +H_{ij},\\
\gamma&=&\Psi^3-\frac{1}{2}\Psi h^{\rm TT}_{ij}h^{\rm TT}_{ij}
+\frac{1}{3}h^{\rm TT}_{ij}h^{\rm TT}_{jk}h^{\rm TT}_{ki},
\end{eqnarray}
being $\chi \equiv \Psi^2 - \frac{1}{2}h_{ij}^{\rm TT}h_{ij}^{\rm TT}$
and $H_{ij}\equiv h^{\rm TT}_{ik}h^{\rm TT}_{jk}-\Psi h^{\rm TT}_{ij}$,
for the inverse and the determinant of the metric, respectively.

\subsection{Equation for the lapse}

Combining relations (\ref{Eq.2.4}) and (\ref{Eq.2.6})
it is straightforward to obtain an elliptic equation for $\alpha$,
\begin{eqnarray}
\gamma^{ii}\alpha_{|j}{}^{j} - \alpha^{|ii} &=&
- \alpha \mbox{R}^{ii}
+\left(\gamma^{mm}\delta_{ik}- 2\gamma^{ik}\right) \alpha \gamma^{-1} \pi^{ij} \pi_{jk}
-\frac{\alpha}{2\gamma}\gamma^{ii}\pi^2
- 2 \gamma^{-1/2} \pi^{im} \beta^i_{~,m}
\nonumber \\\label{alphaeq2}
&+& \frac{\kappa}{2} ~ \alpha \left(\gamma^{il}\gamma^{im} T_{lm}+
\gamma^{ii} \alpha^2T^{00}\right),
\end{eqnarray}
where we have made use of the Hamiltonian constraint (\ref{Eq.2.2}) to get rid of the
Ricci scalar. Note that the last term of the first line is corrected from Ref. \cite{ScGo04}.
The left-hand side of this equation can be given as,
\begin{eqnarray}
\left(\gamma^{jm}\gamma^{ii}-\gamma^{ji}\gamma^{mi}\right)
\left(\alpha_{,jm}-\gamma^{kl}\Gamma_{ljm} \alpha_{,k}\right),
\end{eqnarray}
where, the inverse metric combination can be rewritten making use
of the formulas presented in the previous section,
\begin{equation}
(\gamma^{jm}\gamma^{ii}-\gamma^{ji}\gamma^{mi})=
\frac{2\chi\Psi^2}{\gamma^2}\delta_{jm}+\frac{1}{\gamma^2}\biggl[
\left(\chi+H_{ii}\right)H_{jm}-H_{ij}H_{im}
\biggr].
\end{equation}
Therefore, the equation for the lapse takes the explicit form,
\begin{eqnarray}
\Delta\alpha&=&-\frac{1}{2\chi\Psi^2}\biggl[
\left(\chi+H_{ii}\right)H_{jm}-H_{ij}H_{im}
\biggr]\left(\alpha_{,jm}-\gamma^{kl}\Gamma_{ljm} \alpha_{,k}\right)
+\gamma^{kl}\Gamma_{ljj}\alpha_{,k}
\nonumber\\
&+&\frac{\gamma^2}{2\chi\Psi^2}\biggl[
- \alpha \mbox{R}^{ii}
+\left(\gamma^{mm}\delta_{ik}- 2\gamma^{ik}\right) \alpha \gamma^{-1} \pi^{ij} \pi_{jk}
-\frac{\alpha}{2\gamma}\gamma^{ii}\pi^2
- 2 \gamma^{-1/2} \pi^{im} \beta^i_{~,m}
\nonumber \\\label{alphaeq}
&+& \frac{\kappa}{2} ~ \alpha \left(\gamma^{il}\gamma^{im} T_{lm}+
\gamma^{ii} \alpha^2T^{00}\right)
\biggr].
\end{eqnarray}

The only term that is left to be known in this equation in terms
of metric components is
the noncovariant trace of the Ricci tensor $\mbox{R}^{ii}$.
This tensor can be written in the following way,
\begin{eqnarray}
2 \gamma \mbox{R}_{ij}  &=&
- \chi \biggl [
\Delta h^{\rm TT}_{ij} + \delta_{ij} \Delta\Psi  + \Psi_{,ij} \biggr]
- H_{kl}
(\gamma_{kl,ij}  + \gamma_{ij,kl}
- \gamma_{kj,il} - \gamma_{il,kj} )
\nonumber \\\label{Rij}
&&
+ 2 \gamma \gamma^{kl} \gamma^{np}
(\Gamma_{nil} \Gamma_{pkj} - \Gamma_{nij} \Gamma_{pkl})\,,
\end{eqnarray}
and its trace is then given by,
\begin{eqnarray}
\mbox{R}^{ii}&=&-\frac{2\chi^3}{\gamma^3}\Delta\Psi
- \frac{\chi}{2\gamma^3}\biggl[
2\chi H_{ij}
+H_{im}H_{jm}
\biggr ] (\Delta h^{\rm TT}_{ij}
+\delta_{ij}\Delta\Psi+\Psi_{,ij})
\nonumber\\
&-&\frac{1}{2\gamma}\gamma^{im}\gamma^{jm} H_{kl}
(\gamma_{kl,ij}  + \gamma_{ij,kl}
- \gamma_{kj,il} - \gamma_{il,kj} )
\nonumber \\\label{Rii}
&+&\gamma^{im}\gamma^{jm}\gamma^{kl} \gamma^{np}
(\Gamma_{nil} \Gamma_{pkj} - \Gamma_{nij} \Gamma_{pkl}).
\end{eqnarray}
In this expression one should replace the Laplacian of the conformal
factor that appears in the first term of the right-hand side with
its corresponding Eq. (\ref{psieq}), which will be presented
in the next subsection, in order to enforce a decay rate of
$\sim 1/r^4$ of the right-hand side of Eq. (\ref{alphaeq}).

\subsection{Equation for the conformal factor}

We will use the Hamiltonian constraint (\ref{Eq.2.2}) to solve
for the conformal factor $\Psi$. The Ricci scalar can be written
in the following way,
\begin{equation}
\mbox{R}=(\gamma^{ik}\gamma^{jl}-\gamma^{ij}\gamma^{kl})
(\Psi_{,ij}\delta_{kl}+h^{\rm TT}_{ij,kl})
+\gamma^{kl} \gamma^{np} \gamma^{ij}
(\Gamma_{nil} \Gamma_{pkj} - \Gamma_{nij} \Gamma_{pkl}),
\end{equation}
where the first- and second-order derivatives of metric
components are clearly separated. Expanding it we get
the following more explicit expression,
\begin{eqnarray}\nonumber
\mbox{R}&=& -\frac{2\chi\Psi^2}{\gamma^2}\Delta \Psi -
\frac{\chi}{\gamma^2} H_{ij}\,\Delta h^{\rm TT}_{ij}
-\frac{1}{\gamma^2}\biggl[(\chi+H_{kk})H_{ij}-H_{ik}H_{jk}\biggr]\Psi_{,ij}
\\\nonumber
&+&\frac{1}{\gamma^2}\biggl[H_{ik}H_{jl}-H_{ij}H_{kl}\biggr]h^{\rm TT}_{ij,kl}
+\gamma^{kl} \gamma^{np} \gamma^{ij}
(\Gamma_{nil} \Gamma_{pkj} - \Gamma_{nij} \Gamma_{pkl})\,.
\end{eqnarray}
Now it is straightforward to introduce this expression in the
Hamiltonian constraint obtaining, in this way, an elliptic
equation to solve for the conformal factor,
\begin{eqnarray}\nonumber
\Delta \Psi &=&-
\frac{1}{2\Psi^2} H_{ij}\,\Delta h^{\rm TT}_{ij}
-\frac{1}{2\chi\Psi^2}\biggl[(\chi+H_{kk})H_{ij}-H_{ik}H_{jk}\biggr]\Psi_{,ij}
\\\nonumber
&+&\frac{1}{2\chi\Psi^2}\biggl[H_{ik}H_{jl}-H_{ij}H_{kl}\biggr]h^{\rm TT}_{ij,kl}
+\frac{\gamma^2}{2\chi\Psi^2}\gamma^{kl} \gamma^{np} \gamma^{ij}
(\Gamma_{nil} \Gamma_{pkj} - \Gamma_{nij} \Gamma_{pkl})
\\\label{psieq}
&-&\frac{\gamma}{4\chi\Psi^2}(2\pi^{ij} \pi_{ij}-\pi^2)-\frac{\kappa}{2\chi\Psi^2} \alpha^2\gamma^2 T^{00}\,.
\end{eqnarray}

\subsection{Equation for the longitudinal field momentum}

Because of the gauge condition (\ref{Eq.2.6}), we can perform the
following decomposition for $\pi^{ij}$,
\begin{equation}
\pi^{ij} =  \tilde\pi^{ij} +\pi^{ij}_{\rm TT},
\end{equation}
where the longitudinal part $\tilde\pi^{ij}$ can be given
in terms of a vector field,
\begin{equation}\label{longitudinalpi}
\tilde\pi^{ij} =  \pi^j{}_{,i} + \pi^i{}_{,j}
-\frac{2}{3}\delta_{ij}\pi^m{}_{,m}.
\end{equation}
An elliptic equation for this vector can be
obtained from Eq. (\ref{Eq.2.3}), since
$\pi^{ij}{}_{,j}=\Delta\pi^i +\frac{1}{3}\pi^j{}_{,ij}$,
\begin{equation}\label{pieq}
\Delta\pi^i +\frac{1}{3}\pi^j{}_{,ij}
=
-\gamma^{ik}\Gamma_{kjl}\pi^{jl}
-\frac{\kappa}{2}\gamma^{1/2}\alpha\gamma^{ij}T^0{}_j.
\end{equation}

In summary, we have fixed the eight (out of 16) degrees of freedom in
the initial slice through gauge conditions (\ref{Eq.2.6}) and (\ref{Eq.2.7})
together with the constraint equations, which are implemented by
the elliptic equations for the conformal factor (\ref{psieq})
and for the longitudinal part of the momentum (\ref{pieq}).
Finally, Eqs. (\ref{betaeq}) and (\ref{alphaeq})
for the shift and lapse guarantee that the mentioned gauge
conditions are conserved through the evolution.

The only objects that are then left to be fixed are the transverse
and traceless part of the metric $h_{ij}^{\rm TT}$ and its
conjugate momentum $\pi^{ij}_{\rm TT}$. The four components
of these tensors describe, in this setting, the two physical
degrees of freedom of the theory. This means that they are
freely specifiable on the initial slice but the conditions
imposed on them will restrict the physical problem we are
dealing with.

\section{A 2PN-exact waveless condition: $\pi^{ij}_{\rm TT} \equiv 0$}\label{2PN}

Thus, in order to close the set of equations and obtain
an approximated version of Einstein theory,
we will introduce the following restrictions,
\begin{equation}
\pi^{ij}_{\rm TT}\equiv 0 \quad {\rm and ~ thus},\quad \pi^{ij}_{\rm TT}{}_{,0}\equiv 0.
\end{equation}
This assumption is exact at 2PN order since, in the general
scenario, $\pi^{ij}_{\rm TT}$ is vanishing up to ${\cal O}(c^{-5})$
and it can be understood as the conjugate condition
to that of the CFC case, where $h^{\rm TT}_{ij}$ and its time
derivative are assumed to be vanishing. Therefore, in this approach,
the conjugate momentum $\pi^{ij}$ will be purely
longitudinal and, hence, the vector field $\pi^i$ will contain all the
needed information in order to reconstruct it (\ref{longitudinalpi}).
Thereby now the only missing equation is that for $h^{\rm TT}_{ij}$
and it will be obtained from the TT part of Eq. (\ref{Eq.2.4}).
Let us denote by $A^{ij}$ the right-hand side of that equation,
\begin{equation}\nonumber
A^{ij} \equiv - \alpha \gamma^{1/2} \mbox{R}^{ij} + Y^{ij} +\alpha_{,ij},
\end{equation}
where we have defined,
\begin{eqnarray}\nonumber
Y^{ij}&\equiv &\frac{\alpha}{2\gamma^{1/2}}(2\gamma^{ij}\pi^{kl}\pi_{kl}
-4\pi^{ik} \pi^j{}_k+2\pi\pi^{ij}-\pi^2\gamma^{ij})
\\\nonumber
&+& \gamma^{1/2}(\gamma^{ik}\gamma^{jm}
-\gamma^{ij}\gamma^{km})(\alpha_{,km} -\gamma^{ln}\Gamma_{nkm} \alpha_{,l})
-\alpha_{,ij}
\\
&+& (\pi^{ij}\beta^m)_{,m}
- \pi^{mj} \beta^i{}_{,m} - \pi^{mi} \beta^j{}_{,m}
+ \frac{\kappa}{2} ~ \alpha \gamma^{1/2}(\gamma^{il}\gamma^{jm} T_{lm}
+ \gamma^{ij} \alpha^2 T^{00}),
\end{eqnarray}
and use has been made of the Hamiltonian constraint (\ref{Eq.2.2})
in order to remove the Ricci scalar from this expression.
The term $\alpha_{,ij}$ has been added and subtracted from the
previous definition in order to ensure a decay rate of
$1/r^4$ for $Y^{ij}$ at spacelike infinity.

On the other hand, raising indices in formula (\ref{Rij}),
the contravariant Ricci tensor is given by,
\begin{eqnarray}
\mbox{R}^{ij}=-\frac{\chi^3}{2\gamma^3}\left(\Delta h_{ij}^{\rm TT}+\Psi_{,ij}
+\delta_{ij}\Delta\Psi\right)+X^{ij},
\end{eqnarray}
with
\begin{eqnarray}
X^{ij}&\equiv &-\frac{\chi}{2\gamma^3}
\left[
\Delta h^{\rm TT}_{km} +\delta_{km}\Delta\Psi+\Psi_{,km}
\right]
\bigg\{
\chi\delta_{mi}H_{jk}
+\chi\delta_{kj}H_{im}
+ H_{im}H_{jk}
\bigg\}
\nonumber\\
&-&\frac{1}{2\gamma}\gamma^{ik}\gamma^{jm}
(\gamma_{np,km}  + \gamma_{km,np}
- \gamma_{nm,kp} - \gamma_{kp,nm} )H_{np}
\nonumber\\
&+&\gamma^{ik}\gamma^{jm}
\gamma^{rs}\gamma^{np}(\Gamma_{nks}\Gamma_{prm}-\Gamma_{nkm}\Gamma_{prs}).
\end{eqnarray}

With these definitions at hand, one can write down $A^{ij}$ as,
\begin{eqnarray}
A^{ij}&=&\frac{1}{2}\left(
\frac{\alpha\chi^3}{\gamma^{5/2}}-1\right)
\left(\Delta h_{ij}^{\rm TT}+\Psi_{,ij}
+\delta_{ij}\Delta\Psi\right)
-\alpha\gamma^{1/2}X^{ij} + Y^{ij} +\alpha_{,ij}
\nonumber\\
&+&\frac{1}{2}\left(\Delta h^{\rm TT}_{ij}
+\Psi_{,ij}+\delta_{ij}\Delta\Psi
\right),
\end{eqnarray}
where, again, terms in the second line have been added and
subtracted for future convenience.
Because of the gauge condition (\ref{Eq.2.6}), $A^{ij}$ is
tracefree on shell. Even though, in order to make an eventual numerical
evolution of this system more stable, we will make it
explicitly tracefree. After that we will
need to obtain the transverse part
of the resulting object. In order to do so, we take into account that
full-derivative terms do not contribute to the transverse and
traceless part, thus both $\alpha_{,ij}$ and
$\Psi_{,ij}$ can be eliminated, whereas explicit pure-trace terms
(those proportional to $\delta_{ij}$) can be included in the
trace part $A$. Therefore, the evolution equation for
$\pi^{ij}_{\rm TT}$ (\ref{Eq.2.4}) is now rewritten as,
\begin{equation}\label{pidoteq}
\pi^{ij}_{\rm TT}{}_{,0}=\frac{1}{2}\left(\Delta h^{\rm TT}_{ij}
-B_{\rm TT}^{ij}\right),
\end{equation}
where the source term is given by,
\begin{equation}
B^{ij}_{\rm TT}\equiv
B^{ij}-\frac{1}{3}\delta_{ij}B
-V^j{}_{,i}-V^i{}_{,j}+\frac{2}{3}\delta_{ij}V^l{}_{,l}\,,
\end{equation}
being $B$ the noncovariant trace of $B^{ij}$, i.e., $B\equiv B^{ii}$,
and
\begin{equation}\label{defB}
B^{ij}\equiv\left(1-
\frac{\alpha\chi^3}{\gamma^{5/2}}\right)
\left(\Delta h_{ij}^{\rm TT}+\Psi_{,ij}\right)
+2\alpha\gamma^{1/2}X^{ij} - 2Y^{ij}.
\end{equation}
Finally, the vector field $V^i$ encodes the longitudinal part
of $B^{ij}$ and hence obeys the following equation,
\begin{equation}\label{Veq}
\Delta V^i +\frac{1}{3}V^j{}_{,ji}=B^{ij}{}_{,j}-\frac{1}{3}B_{,i}\,.
\end{equation}

Imposing now our assumption ($\pi_{\rm TT}^{ij} \equiv \pi_{\rm TT}^{ij}{}_{,0} \equiv 0$)
on Eq. (\ref{pidoteq}), it is straightforward to get
an elliptic equation for $h_{ij}^{\rm TT}$,
\begin{equation}\label{heq}
\Delta h_{ij}^{\rm TT}=B^{ij}_{\rm TT}.
\end{equation}
This assumption must also apply on other elliptic equations
presented in last section, obtaining in this way a closed system of
equations which does not contain $\pi_{\rm TT}^{ij}$ .
Because of the terms that have been added and subtracted
previously, the source $B^{ij}_{\rm TT}$ decays properly ($\sim 1/r^4$)
for an eventual numerical resolution of this equation.
In particular, note that the terms contained inside the first brackets of the
definition of $B^{ij}$ (\ref{defB}) have been constructed such that
their combination
decays as $\sim 1/r$ at asymptotic infinity and, moreover,
it is of order ${\cal O}(c^{-2})$, as can be verified using
the expressions (\ref{pnseries}) below.

In summary, the gravitational metric functions are encoded in the objects
$\{\beta_i, \alpha, \Psi, h_{ij}^{\rm TT} \}$ which are connected
with the auxiliary variables $\{\pi^i, V^i\}$
and the corresponding elliptic equations are,
once imposed the condition $\pi^{ij}_{\rm TT}=0$
in all of them, (\ref{betaeq}),
(\ref{alphaeq}), (\ref{psieq}), (\ref{heq}), and (\ref{pieq}),
(\ref{Veq}), respectively.

This set of equations has two main applications.
On the one hand, it can be used to obtain initial data,
such that $\pi^{ij}_{\rm TT}=\pi^{ij}_{\rm TT}{}_{,0}=0$,
for a subsequent numerical evolution of the full Einstein
equations. On the other hand, it could be solved as a
self-contained theory of gravitation, which would not
have waves, but would approximately describe the motion
of the astrophysical bodies.

Regarding this last application,
we would like to end this section by analyzing the
accuracy of our theory, that is, which is the error one
would commit when solving our approximated set of equations
with respect to solving the
full Einstein equations. Since there is no approximation
in the equations for the rest of the objects, the only
error is introduced in the system via the assumption
on $\pi^{ij}_{\rm TT}$. Looking at their
respective equations, and recalling their leading order
post-Newtonian series, namely,
\begin{eqnarray}\label{pnseries}
&\alpha = 1 + {\cal O}\left(c^{-2}\right),
\quad\beta_i = {\cal O}\left(c^{-3}\right),
\quad\pi^{ij} ={\cal O}\left(c^{-3}\right),\quad
\nonumber\\
&\Psi = 1 + {\cal O}\left(c^{-2}\right),
\quad h^{\rm TT}_{ij} ={\cal O}\left(c^{-4}\right),
\end{eqnarray}
one can check that if $\pi^{ij}_{\rm TT}{}_{(n)}$ were computed properly,
where the subscript $(n)$ denotes the coefficient of $1/c^{n}$ in
the corresponding post-Newtonian expansion, the metric components
$\{\alpha_{(n+3)}, \beta^i_{(n+2)}, \gamma_{ij}{}_{(n+1)}\}$
would be obtained correctly. In fact the conformal factor is
calculated up to $\Psi_{(n+3)}$, otherwise we could not compute
the lapse up to the mentioned order.
As we have already commented,
the assumption we have proposed is correct up to 2PN order,
i.e., we are computing $\pi^{ij}_{(3)}$ correctly thus,
by using the proposed scheme,
$\{\alpha_{(6)}, \beta^i_{(5)}, \gamma_{ij}{}_{(4)}\}$
can be properly determined.

\section{A 4PN-exact approximation}\label{4PN}

In this section we will generalize the scheme we have presented in
the previous section, which will give rise to a formalism that
agrees with General Relativity up to 4PN order. Contrary
to our 2PN scheme, this more general approach will indeed
contain gravitational waves since the main equation to
obtain $h^{\rm TT}_{ij}$ will be of hyperbolic nature.

Let us write the evolution equation for the spatial metric (\ref{Eq.2.5})
in the form
\be
\gamma_{ij,0} = C_{ij}\,,
\ee
where 
\be
C_{ij} \equiv  \alpha \gamma^{-1/2}(2\pi_{ij} - \gamma_{ij}\pi) + \beta_{i|j}
+ \beta_{j|i}.
\ee
Using the gauge conditions on $\gamma_{ij}$, that are given by Eq.
(\ref{Eq.2.7}), we obtain the evolution equation for the
transverse and traceless part of the metric,
\be
h^{\rm TT}_{ij,0}  =  C_{ij} - \frac{1}{3}C_{ll}\delta_{ij}\,,
\ee
 where 
$ C_{ii} = 3\eta (C^{ii} 
- \gamma^{ij} \gamma^{ik} h^{\rm TT}_{jk,0}) $ with  $C^{ii}=
- \alpha \gamma ^{-1/2} \gamma^{ii} \pi + 2 \beta^{i|i} $ and
$\eta\equiv (\gamma^{ij}\gamma^{ij})^{-1}$.
This equation can then be rearranged so that
$\pi^{ij}$ follows in terms of $h^{\rm TT}_{ij,0}$ in the form
\bea
M^{ij}{}_{kl} \,\pi^{kl} &=& 
\frac{\gamma^{1/2}}{2\alpha}
\biggl[ 2 \eta\,\beta^{m|m}  \gamma^{in} \gamma^{jn} 
-\beta^{i|j} - \beta^{j|i} 
\nonumber\\[1ex]
&+&  (\gamma^{im} \gamma^{jn} - \eta\,\gamma^{ik} \gamma^{jk}
\gamma^{lm} \gamma^{ln})h^{\rm TT}_{mn,0}\biggr],
\label{Eq.3.11}
\eea
where the matrix $M^{ij}{}_{kl}$ is given by
\be
M^{ij}{}_{kl} =  \frac{1}{2} \biggl[\delta_{il}\delta_{jk}   + \delta_{ik}\delta_{jl} -  
(\gamma^{ij} - \eta\,\gamma^{in} \gamma^{jn}  \gamma^{mm} )
h^{\rm TT}_{kl} \biggr].
\ee
Remarkably, the matrix $M^{ij}{}_{kl}$ deviates from the unit identity matrix in the
quadratic order of $h^{\rm TT}_{ij}$ only \cite{ScGo04}.
Taking into account that  $h^{\rm TT}_{ij}$ is of 2PN order (\ref{pnseries}),
neglecting the quadratic terms of $h^{\rm TT}_{ij}$
in $M^{ij}{}_{kl}$ is a truncation at 5PN order only. Thus,
we can write down
\bea\label{fullpibeta}
\pi^{ij} &=& 
\frac{\gamma^{1/2}}{2\alpha} \biggl[ 2\eta\, \beta^{m|m}  \gamma^{in} \gamma^{jn}
-\beta^{i|j} - \beta^{j|i}                \nonumber\\[1ex]
&+&  (\gamma^{im} \gamma^{jn} -
\eta\,\gamma^{ir} \gamma^{jr} \gamma^{sm} \gamma^{sn})h^{\rm TT}_{mn,0}\biggr]
+{\cal O}(c^{-11})\,,
\eea
which is a 4.5PN-exact relation. Keeping only the leading order of the
time derivative of $h^{\rm TT}_{ij}$ and calculating the transverse
and tracefree part of this relation, we obtain,
\be
\pi^{ij}_{\rm TT}=\frac{1}{2} h^{\rm TT}_{ij,0} +D^{ij} -D^{ij}_{\rm L}
+{\cal O}(c^{-9}),
\label{defpi}
\ee
where
\bea
D^{ij} \equiv - \frac{\gamma^{1/2}} {2\alpha} \biggl(\beta^{i|j} + \beta^{j|i}
- 2 \eta\, \beta^{m|m}  \gamma^{in} \gamma^{jn} \biggr)\,,
\eea
and 
\be
D^{ij}_{\rm L} = W^i_{~,j}  + W^j_{~,i} - \frac{2}{3} W^l_{~,l}~\delta_{ij},
\ee
hold with 
\be\label{Weq}
 \Delta W^i  + \frac{1}{3}W^j_{~,ji} =  D^{ij}_{~~,j}.
\ee
Note that the decay rate of the right-hand side of the last equation
is $1/r^3$ at asymptotic infinity. As we have commented, this could
give problems when solving this equation numerically. Even though,
since the term in question is a full divergence, the vector $W^i$
can be redefined including in it the terms that goes like
$\sim 1/r^3$, so that the source for the equation for the new
vector decays properly.

In order to check that Eq. (\ref{defpi}) is valid up to
${\cal O}(c^{-9})$ one has to take into account the fact that
$\alpha^{2n}\Psi^n=1+{\cal O}(c^{-4})$, as can
be verified with the first terms of their PN expansions:
$\Psi=1+2 U/c^2$ and $\alpha=1-U/c^2$,
$U$ being the Newtonian potential.

The (nonapproximated) evolution equation for $\pi^{ij}_{\rm TT}$ (\ref{pidoteq})
and the (approximated) one for $h_{ij}^{\rm TT}$ (\ref{defpi}), 
\be
 h^{\rm TT}_{ij,0}= 2\pi^{ij}_{\rm TT} - 2\left(D^{ij} -D^{ij}_{\rm L}\,\right)
+{\cal O}(c^{-9}),
\label{defpibis}
\ee
define our system of truncated evolution equations of the Einstein
theory put into canonical form. Combining the mentioned two equations,
and thus making an implicit change from a Hamiltonian to a
Routhian framework (see, e.g., \cite{JS98}), results into a second-order hyperbolic equation
for $h^{\rm TT}_{ij}$,
\be\label{waveeq}
-h^{\rm TT}_{ij,00} + \Delta h^{\rm TT}_{ij} = 2 \left(D^{ij} - D^{ij}_{\rm
L}\,\right)_{,0}  +B^{ij}_{TT} +{\cal O}(c^{-10}),
\ee
which describes the propagation of the two gravitational degrees
of freedom. Defining the source
$S^{\rm TT}_{ij}\equiv2\left(D^{ij} - D^{ij}_{\rm L}\,\right)_{,0}  +B^{ij}_{TT}$,
the no-incoming radiation formal solution of 
this equation is given by the standard retarded integral
(see below for an iterative solution),
\begin{equation}\label{retarded}
h^{\rm TT}_{ij} (t, \vec{x})= - \frac{1}{4\pi}\int \frac{d^3y}{ | \vec{x}- \vec{y}|}S^{\rm TT}_{ij}(t_{\rm ret},  \vec{y})\,,
\end{equation}
with the retarded time $t_{\rm ret} = t -  | \vec{x}- \vec{y}|/c$; for another
representation of the retarded solution see, e.g., \cite{Kanwal}.

In summary, the system of equations to be solved is composed
by the elliptic Eqs. (\ref{betaeq}), (\ref{alphaeq}), (\ref{psieq}),
(\ref{pieq}), (\ref{Veq}) and the wave Eq. (\ref{waveeq})
for $h_{ij}^{\rm TT}$. This set of equations can be understood
as a truncated version of Einstein equations so, by assuming
certain initial data for the hyperbolic equation, they provide
an approximated solution of General Relativity.
On the other hand, they can also be used to obtain initial
data for a subsequent evolution via full Einstein equations.
Here we propose an iterative scheme to construct such initial
data for the metric coefficients. As a first step, one imposes
the assumptions $\pi^{ij}_{\rm TT}=0$ and $h^{\rm TT}_{ij}=0$ in all
elliptic equations and solves them including the equations for the
matter variables for all past times. Then, with
this information, one can solve the hyperbolic Eq. (\ref{waveeq})
for $h_{ij}^{\rm TT}$ via the retarded integral
(and hence without any ``instant-of-time'' initial conditions)
dropping the time derivatives
of the right-hand side since they are of ${\cal O}(c^{-6})$.
In a second step, by solving again the elliptic system with the previously
computed $h_{ij}^{\rm TT}$, and still $\pi^{\rm TT}_{ij}=0$,
a 2.5PN-exact solution $\{\alpha_{(7)}, \beta^i_{(6)}, \gamma_{ij}{}_{(5)}\}$
is obtained. Note that this also 2PN-exact solution is
essentially different from that obtained by solving
the scheme proposed in Sec. \ref{2PN}.
This 2.5PN-exact solution can now be used to compute
$\pi^{ij}_{TT(5)}$ properly through its truncated definition (\ref{defpi}).
With this information at hand, we can calculate the sources
(all the right-hand sides) of our system of equations
to a PN-level of precision higher than in the second step. Therefore,
by solving this system, a 3PN (and thus also 3.5PN)-exact solution can be obtained,
$\{\alpha_{(8)}, \beta^i_{(7)}, \gamma_{ij}{}_{(6)}\}$.
Repeating again this process,
the metric components can be computed up to 4PN-exact order,
$\{\alpha_{(10)}, \beta^i_{(9)}, \gamma_{ij}{}_{(8)}\}$.
The limit of this procedure is given by the order up to which
Eq. (\ref{defpi}) is valid. This iterative scheme generalizes a
procedure which is under development in \cite{KT07,KT09}.

At this stage some remarks are in order.

In this second-order formalism, $\pi^{ij}_{\rm TT}$
has to be replaced by $h^{\rm TT}_{ij~,0}$ in all our equations
using its truncated definition (\ref{defpi}).
This time derivative, as well as the first-order time derivatives
on the right-hand side of the wave Eq. (\ref{waveeq}),
have to be evaluated applying matter evolution equations,
e.g., those that will be presented in Sec. \ref{Matter}.

Regarding the decay rates at asymptotic infinity,
from Eq. (51) in Ref. \cite{JS98}, it follows that at 3PN level the
relation $D^{ij} - D^{ij}_{\rm L} = (\Psi_{(2)}{\tilde \pi}^{ij}_{(3)})^{\rm TT}$
is valid. Since for large radii $\tilde \pi^{ij}_{(3)}$ includes the total linear momentum,
this implies that the difference $(D^{ij} - D^{ij}_{\rm L})_{,0}$,
that is present in (\ref{waveeq}), decays as $\sim 1/r^4$ at asymptotic
infinity due to conservation of linear momentum. With respect to the
implicitly present $h^{\rm TT}_{ij~,0}$ in many sources,
the conservative motion which enters into the source terms of
the hyperbolic equation does not allow
for a solution with time derivative decaying at least as $1/r^2$. Only
at higher PN orders, when radiation damping is entering, the matter motion
produces decaying wave amplitudes towards spacelike infinity. In order
to achieve a proper decay on our level
of approximation a technical trick, similar to adiabatic damping,
may be introduced as follows. All partial time and space
derivatives $\partial_{\mu}$ (with $\mu = 0,i$) of $h_{ij}^{\rm TT}$,
that appear in
source terms, should be replaced through $\mbox{exp}(-\epsilon r/r_0) \partial_{\mu}$,
where $r$ is the radial
coordinate originating from the center of energy, $r_0$ a typical
radial extension of the matter distribution, and $\epsilon$ a small
positive constant, $0 <  \epsilon << 1$. At the end of the day,
the final result should depend neither on $\epsilon$ nor on $r_0$.
For a related treatment of retarded solutions see, e.g., \cite{Blanchet}.
In conclusion, applying this procedure, through 3PN order,
and even 3.5PN, all source terms in the elliptic and hyperbolic equations
to be solved decay nicely at spacelike infinity. On the other hand,
at the 4PN level tail terms will have to be controlled as
it is evident from the Introduction.

Finally, the Eq.
(\ref{fullpibeta}) shows that for stationary configurations, like for Kerr
black holes, where $h^{\rm TT}_{ij~,0}$ vanishes, our approach is 5PN exact,
that is, it provides $\{\alpha_{(12)}, \beta^i_{(11)}, \gamma_{ij}{}_{(10)}\}$
accurately.
This context suggests another 2PN-exact waveless approach through 
dropping $h^{\rm TT}_{ij~,0}$ in Eq. (\ref{fullpibeta}) and also all terms in
Eq. (\ref{waveeq}) with time derivatives, which is equivalent
to the assumption $\pi^{ij}_{\rm TT}{}_{,0}=0$ in Eq. (\ref{pidoteq}).
In this way, a quasistationary
approximation scheme is achieved which is 5PN exact for stationary configurations
and covers both the 1PN-exact CFC and the 2PN-exact CFC+
approach (see next section). This latter approach is closest
to \cite{ScGo04}, where $\pi^{ij}_{\rm TT,0}$ is also put equal to zero
in Eq. (\ref{pidoteq}) but Eq. (\ref{Eq.3.11})
is kept exact.

\section{Relation to 2PN-exact CFC+ approach}\label{CFC}

In this section we want to reduce our scheme to CFC+ order
in order to compare with previous results in the literature.
As has already been explained,
the CFC case is nothing but allowing $h^{\rm TT}_{ij}$ to
be vanishing. This is the only approximation that is done
in this approach, so all the equations that we will present
in this section will be fully correct in the conformal flat
case just by removing terms with $h^{\rm TT}_{ij}$. On
the other hand, the CFC+ approach considers also linear
terms in $h^{\rm TT}_{ij}$ hence, in this case, the
following definitions apply,
\begin{equation}
\gamma^{ij}=\Psi^{-1}\delta_{ij}-h^{\rm TT}_{ij},\quad
\gamma=\Psi^3,\quad\chi=\Psi^2,
\quad {\rm and}\quad \eta=\frac{\Psi^2}{3}.
\end{equation}
The time derivative of $h^{\rm TT}_{ij}$ in Eq. (\ref{fullpibeta})
can also be neglected, including however a nonvanishing $\pi^{ij}_{\rm TT}$
which makes it a 2PN-exact approach different from Sec. IV,
so for $\pi^{ij}$ only
\bea \label{pibeta}
\pi^{ij} &=& -
\frac{\gamma^{1/2}}{2\alpha} \biggl(\beta^{i|j} + \beta^{j|i}
 - 2\eta\, \beta^{m|m}  \gamma^{in} \gamma^{jn}\biggr) \nonumber\\ 
&=&-\frac{\Psi^{1/2}}{2\alpha}\left(\beta^i{}_{,j}+\beta^j{}_{,i}
-\frac{2}{3}\delta_{ij} \beta^k{}_{,k}\right),
\eea
is needed. In fact, this equation is only used in the CFC case
since in the CFC+ scheme, it is only valid up to 1PN order $c^{-3}$
counting in terms of integer PN orders.

At this linearized order, our equation for the shift,
\begin{equation}
\Delta\beta_i +\frac{1}{3}\beta_{j,ji}=
-\frac{2}{3}(\gamma^{jk}\beta_j \Gamma_{kll})_{,i}
+2 (\gamma^{jk}\beta_j \Gamma_{kil})_{,l}- 2(\alpha \gamma^{-1/2} \pi_{ij})_{,j},
\end{equation}
is trivially obeyed when taking into account last relation (\ref{pibeta}).
Hence, in order to obtain a meaningful equation for the shift, we introduce
relation (\ref{pibeta}) in the momentum constraint and obtain,
\begin{equation}
\Delta\beta^i +\frac{1}{3}\beta^j{}_{,ji}=-2\Psi
\left(\frac{\alpha}{\Psi^{3/2}}\right)_{,j}\pi^{ij}
+\Psi\kappa\alpha^2 \gamma^{ij}T^0{}_j.
\end{equation}

In the equations for the lapse (\ref{alphaeq})
and the conformal factor (\ref{psieq}), the
terms that survive at this order are given by,
\begin{eqnarray}
\Delta\alpha &=& \frac{1}{2} h^{\rm TT}_{ij}\alpha_{,ij}
+\gamma^{kl}\Gamma_{ljj} \alpha_{,k}
+\frac{\gamma^2}{2\chi\Psi^2}\biggl[
- \alpha \mbox{R}^{ii}+\frac{\alpha}{\gamma\Psi}\pi^{ij}\pi_{ij}
-2\gamma^{-1/2}\pi^{im}\beta^i{}_{,m}\nonumber\\
&+& \frac{\kappa}{2} ~ \alpha (\gamma^{il}\gamma^{im} T_{lm}+
\gamma^{ii} \alpha^2T^{00})
\biggr],\\
\Delta\Psi  &=&\frac{\gamma^2}{2\chi\Psi^2}\gamma^{kl} \gamma^{np} \gamma^{ij}
(\Gamma_{nil} \Gamma_{pkj} - \Gamma_{nij} \Gamma_{pkl})
-\frac{\gamma}{2\chi\Psi^2}\pi^{ij}\pi_{ij}
-\frac{\kappa}{2\chi\Psi^2}\gamma^2\alpha^2T^{00},
\end{eqnarray}
which can be written in a simpler way as
\begin{eqnarray}
\Delta\alpha &=&
-\frac{1}{2\Psi}\Psi_{,i} \alpha_{,i}
-\Psi^{1/2}\pi^{ij}\beta^i{}_{,j}
+\frac{\kappa\alpha}{4} (T_{ii}+
\Psi \alpha^2T^{00})
+\frac{1}{2} h^{\rm TT}_{ij}\alpha_{,ij}
-\frac{1}{4} h^{\rm TT}_{ij}\Psi_{,ij},\\
\Delta\Psi  &=& \frac{3}{4\Psi}\Psi_{,i}\Psi_{,i}
-\frac{1}{2\Psi}\pi^{ij}\pi_{ij}-\frac{\kappa}{2}\Psi^2\alpha^2T^{00}.
\end{eqnarray}
Here, we have used the fact that the noncovariant trace of the Ricci
tensor at this order is given just by,
\begin{equation}
\mbox{R}^{ii}=\frac{\kappa\alpha^2}{\Psi}T^{00}+\frac{1}{\Psi^4}\pi^{ij}\pi_{ij}
+\frac{1}{2}h^{\rm TT}_{ij}\Psi_{,ij}.
\end{equation}

Finally, the set of CFC+ equations is closed by the elliptic equation for
the transverse and traceless part of the spatial metric,
\begin{eqnarray}\label{htteqcfc}
\Delta h_{ij}^{\rm TT}&=& B_{ij}-\frac{1}{3}\delta_{ij}B-V^j{}_{,i}
-V^i{}_{,j}+\frac{2}{3}\delta_{ij}V^m{}_{,m},\\
\Delta V^i +\frac{1}{3}V^j{}_{,ji}&=& B_{ij,j} -\frac{1}{3}B_{,i}\,,
\end{eqnarray}
where $B=B_{ii}$ and now the following definition for $B_{ij}$ should be applied,
\begin{equation}
B_{ij}=\Psi_{,i}\alpha_{,j}+\Psi_{,j}\alpha_{,i}-\kappa T_{ij}.
\end{equation}
In order to arrive to this expression, we have integrated by parts
terms with second derivatives of the lapse and the conformal factor,
taking into account that the full-derivative terms are eliminated when
computing the transverse and tracefree part, as it is done in the
right-hand side of Eq. (\ref{htteqcfc}).
These CFC+ equations have been compared to those presented in
Ref. \cite{CD05} and obtained exact agreement. For such
comparison, one has to consider that, as already shown
in the previous section, $\Psi=1+2 U/c^2$ and
$\alpha=1-U/c^2$ for the 2PN terms beyond CFC.

In conclusion, we have shown that our proposed 4PN-exact
formalism covers naturally both CFC and CFC+ approximations.
Regarding the 2PN-exact approach presented in Sec. \ref{2PN},
it also coincides with the CFC and CFC+ cases up to the mentioned
PN-level. The key difference between the CFC+ scheme
and the one of Sec. \ref{2PN} is that whereas we have kept all the nonlinear
terms present in all the equations, in CFC+ approach
one just keeps those terms that are of the corresponding (2PN)
order. Even though, our 2PN elliptic framework of  Sec. \ref{2PN} can not
``fully'' cover the CFC case since we have assumed that
$\pi^{ij}_{\rm TT}=0$ whereas in the standard CFC scheme
the definition (\ref{pibeta}) is used. Only under spherical symmetry
conditions both $h_{ij}^{\rm TT}$ and $\pi^{ij}_{\rm TT}$ vanish.

\section{A specific matter model}\label{Matter}

In this section we specify the matter model to the particular case of
a barotropic perfect fluid [in this section, $c=1$].
Therefore, the stress-energy tensor will be given in terms of the fluid
four-velocity $u^\mu$, pressure $p$, proper mass
density $\rho$, and specific enthalpy $h$ in the following way,
\begin{equation}
T^{\mu\nu}= \rho (1 + h)u^\mu u^\nu +p g^{\mu\nu},
\end{equation}
where the conservation law $\nabla_{\mu} (\rho u^{\mu})=0$ holds with
$\nabla_{\mu}$ the four-dimensional covariant derivative.

We need the following components of the stress-energy tensor density $(-g)^{1/2} T^{\mu\nu}$
that appear in the equations,
\begin{eqnarray}
\alpha \gamma^{1/2} T^{00}&=& \rho_*(1 + h)u^t -\frac{\gamma^{1/2}p}{\alpha},\\
\alpha \gamma^{1/2} T^0{}_j&=& \rho_*(1+h) u_j,\\
\alpha \gamma^{1/2} T_{ij}&=& \rho_*(1 + h) u_i\frac{u_j}{u^t} + \alpha \gamma^{1/2} p\gamma_{ij},
\end{eqnarray}
where the four-velocity is decomposed as $u^{\mu}=(u^t,u^i),
u_{\mu}=(u_t,u_i)$ and the definition $\rho_* = \alpha
\gamma^{1/2} u^t \rho$ has been made. From the normalization
of the four-velocity $u^\mu u_\mu=-1$, it is easy to see that
$u^t$ is given by $u^t = (1 + \gamma^{ij}u_iu_j)^{1/2}/\alpha$.

In the case of a barotropic perfect fluid, where $dp= \rho~ dh$ holds,
the equation of state $p=p(\rho)$ and the conservation of stress-energy
tensor,
\begin{equation}
\nabla_{\mu} T^{\mu\nu}=0,
\end{equation}
give rise to all the equations of motion for the matter. The independent equations of
motions can be cast into the form \cite{H85,BDS90},
\begin{equation}
\partial_t \rho_*= - \partial_i(\rho_* v^i),
\end{equation}
\begin{equation}
\partial_t P_i  = - \partial_j(P_i v^j) - \partial_i(\alpha
\gamma^{1/2}p) +
\frac{\alpha}{2}\gamma^{1/2} T^{\mu\nu}\partial_ig_{\mu\nu},
\end{equation}
where $v^i \equiv u^i/u^t$ and $P_i \equiv \rho_*(1+h)u_i$. Introducing
$w_i \equiv P_i/\rho_*$, i.e., $w_i = (1+h) u_i$, the latter equation of
motion can be written    
\begin{equation}
\partial_t w_i  = - v^j\partial_j w_i - \frac{1}{\rho_*} \partial_i(\alpha
\gamma^{1/2}p) + \frac{\alpha\gamma^{1/2}T^{\mu\nu}}{2\rho_*}\partial_ig_{\mu\nu}.
\end{equation}
With the aid of the relation 
\begin{equation}
v^i = \frac{\alpha\gamma^{ij}w_j}{[(1+h)^2 + \gamma^{ij}w_iw_j]^{1/2}} - \beta^i, 
\end{equation}
all matter variables can be reduced to the independent ones $(\rho_*,
w_i)$ or $(\rho_*, P_i)$.

\section{Conclusions}\label{Conclusions}

In this paper we have developed a 4PN-exact approximation to the field
equations of General Relativity which turns out to be fully exact in the conformal flat
case and 5PN exact for stationary configurations. The elliptic equations for
the lapse $\alpha$, the shift $\beta^i$, and the conformal factor
$\Psi$ are exact (just the well-known equations from the ADM formalism
but in some more explicit form):
Eqs. (\ref{alphaeq}), (\ref{betaeq}), and (\ref{psieq}), respectively.
Only the wave equation for $h^{\rm TT}_{ij}$ (\ref{waveeq})
and the definition for its conjugate momentum $\pi^{ij}_{\rm TT}$ (\ref{defpi})
are approximated. In order to obtain the mentioned hyperbolic equation,
an implicit transition from a Hamiltonian to a Routhian framework,
regarding the independent gravitational degrees of freedom,
has taken place \cite{JS98}. This equation
has to be solved under the condition of no incoming
radiation. The construction of various transverse-traceless objects resulted in several
auxiliary vectors $\pi^i$, $V^i$, and $W^i$ obeying the Eqs.
(\ref{pieq}), (\ref{Veq}), and (\ref{Weq}), respectively.

Our proposal is to solve the commented set of equations iteratively.
In particular, the first step is to solve the elliptic equations and the
equations of motion of the matter for all times prior to the initial
value slice imposing $h^{ij}_{\rm TT}\equiv 0$ and
$\pi^{\rm TT}_{ij}\equiv 0$. Then, using this result, the
hyperbolic equation for $h^{ij}_{\rm TT}$ as well as the
elliptic equations again are being solved which gives rise to a
2[2.5]PN-exact approximation to the Einstein field equations.
The obtained 2PN-exact solution can be used to correctly compute
$\pi^{ij}_{TT}$ through its truncated definition (\ref{defpi})
up to 3PN order. With this information at hand, the sources
of the mentioned system of equations can be calculated
to one PN-order of precision higher. This fact permits,
by solving again the system, to obtain a 3[3.5]PN-exact solution.
This process can be repeated up to obtain a 4[4.5]PN-exact solution.
The sources have been manipulated so that they decay
fast enough ($\sim 1/r^4$) at asymptotic infinity. 
Therefore the system of equations we present is
well suited for an eventual numerical evolution. At any instant of
time, the obtained solutions can be used as initial data to be
evolved according to the full Einstein equations or, the solutions can
be treated as approximate solutions of the full Einstein equations
throughout all times.  

The presented approach covers recently developed and explored
approaches including outgoing radiation \cite{TB03,KT07,KT09,JY09}.
For binary compact objects, metric coefficients for the
coordinate conditions of the present paper are known 
in closed-analytic form through 2[2.5]PN order in the near zone
\cite{OO74,OK75,S85,JS98}. Those have already been used for further iterations
in, e.g., \cite{KT07,KT09}.

Finally, we have also analyzed the 2PN-exact waveless approximation to General
Relativity that is obtained by imposing the assumptions
$\pi^{ij}_{\rm TT}=\pi^{ij}_{\rm TT}{}_{,0}=0$. These conditions are quite
interesting since they can be considered as the conjugate assumptions
to the well-known conformal flat conditions ($h_{ij}^{\rm TT}=h_{ij,0}^{\rm TT}=0$).
A maximum elliptic system without any partial time derivatives of the gravitational
field variables that one can obtain within General Relativity
is dropping $h_{ij,0}^{\rm TT}$  in Eq. (\ref{Eq.3.11}) and  $\pi^{ij}_{\rm
TT}{}_{,0}$ in Eq. (\ref{pidoteq}). Hereof, the simplified waveless approach suggested
in Sec. \ref{4PN} results by just dropping $h_{ij,0}^{\rm TT}$ in Eq. (\ref{defpi}).
A further truncation is obtained by putting $\pi^{ij}_{\rm TT} \equiv 0$ which is the
waveless approximation of Sec. \ref{2PN}. All these waveless
approximations are 2PN exact where the latter one is the simplest of them.  
In order to obtain more accurate truncated versions of General
Relativity, hyperbolic equations should be considered, as it is done in
the above commented 4PN-exact approach.

\begin{acknowledgments}
G.S. thanks A. Gopakumar for useful discussions. This work is supported by the Deutsche
Forschungsgemeinschaft (DFG) through SFB/TR 7 ``Gravitational Wave Astronomy''.
D.B. is also partially funded by the Spanish MICINN Project
FIS2008-06078-C03-03.

\end{acknowledgments}

\end{document}